\documentclass[
  twoside,
  twocolumn,
  aps,
  prm,
  floatfix,
  citeautoscript,
  superscriptaddress,
]{revtex4-2}
\usepackage[english]{babel}
\usepackage[T1]{fontenc}
\usepackage{upgreek}
\usepackage{textcomp} 				
\usepackage[dvips]{graphicx}		
\usepackage{color}					
\usepackage{amssymb,amsmath}		
\usepackage{siunitx}
\DeclareSIUnit{\liter}{l}
\usepackage[colorlinks=true]{hyperref}
\hypersetup{
    colorlinks=true,        
    linkcolor=black,      
    citecolor=blue,        
    filecolor=blue,      
    urlcolor=blue        
}

\subequations
\usepackage{dblfloatfix}

\renewcommand{\paragraph}[1]{{\bf #1}}

\begin{document}

\title{Band gap tuning by structural phase transition in Sm-substituted BiFeO$_3$ powders} 

\author{Christina Hill}
\affiliation{Department of Physics and Materials Science, University of Luxembourg, 41 rue du Brill, 4422 Belvaux, Luxembourg}
\affiliation{Inter-institutional Research Group Uni.lu–LIST on Ferroic Materials, 41 rue du Brill, 4422 Belvaux, Luxembourg}
\author{Michele Melchiorre}
\affiliation{Department of Physics and Materials Science, University of Luxembourg, 41 rue du Brill, 4422 Belvaux, Luxembourg}
\author{Cosme Milesi-Brault}
\author{Pascale Gemeiner}
\author{Fabienne Karolak}
\author{Christine Bogicevic}
\author{Brahim Dkhil}
\affiliation{Université Paris-Saclay, CNRS, CentraleSupélec, Laboratoire SPMS, 91190 Gif-sur-Yvette, France}
\author{Ingrid Ca\~nero-Infante}
\affiliation{Université de Lyon, Institut des Nanotechnologies de Lyon, CNRS UMR 5270 ECL INSA UCBL CPE, 1 rue Enrico Fermi, F-69621 Villeurbanne, France}
\author{Mael Guennou}
\affiliation{Department of Physics and Materials Science, University of Luxembourg, 41 rue du Brill, 4422 Belvaux, Luxembourg}
\affiliation{Inter-institutional Research Group Uni.lu–LIST on Ferroic Materials, 41 rue du Brill, 4422 Belvaux, Luxembourg}

\begin{abstract}

The substitution of bismuth by samarium in BiFeO$_3$ is known to induce a structural phase transition from the polar phase to a non-polar phase, with a possible antiferroelectric intermediate structure. In this paper, we investigate the impact of this phase change on the optical properties. The optical band gap was measured by diffuse reflectance as a function of temperature for several samarium concentrations across the structural phase transition. We found that the optical band gap for each of the pure phases varies linearly with temperature and that the phase transitions are revealed by smooth transitions between those linear regimes. This allows us to quantify the contribution of the structural change in the optical absorption. We find that a difference in optical band gap of about $\approx\SI{130}{\meV}$ can be attributed to the phase change. We anticipate that the same change could be obtained by applying an electric field in an antiferroelectric composition. 

\end{abstract}

\maketitle


\section*{Introduction}

Bismuth ferrite BiFeO$_3$ (BFO) is a pivotal compound in materials science as a model multiferroic crystal combining ferroelectric properties at room temperature and complex magnetic order. Due to a band gap energy in the visible light range at $\approx\SI{2.4}{\eV}$, BFO has attracted considerable attention for potential use in photocatalysis~\cite{ irfan_critical_2019, supriya_recent_2023, amdouni_bifeo3_2023} and photovoltaic~\cite{choi_switchable_2009, yang_photovoltaic_2009, lotey_gd-doped_2013, tiwari_solution_2015, you_enhancing_2018} applications. Pure BFO crystallizes in a rhombohedral structure with space group $R3c$ and undergoes a first-order phase transition to a paraelectric phase at \SI{1100}{\K}. The substitution of bismuth by a rare earth element $R$ has attracted attention for a long time as a way to tune the structural~\cite{Arnold2015} and optical properties of BFO~\cite{Irfan2017,mumtaz_chemical_2021, gholizadeh_effect_2024}. At sufficiently high $R$ concentrations, the compound stabilizes in an orthorhombic non-polar $Pnma$ crystal structure that is also the structure of the pure rare-earth orthoferrites $R$FeO$_3$ and of the high-temperature paraelectric phase of BFO. The transition between the polar $R3c$ phase (at low $x$) and the non-polar $Pnma$ phase (at high $x$) occurs not only via phase coexistence but also, in a narrow composition range, of intermediate bridging phases usually characterized by non-polar structures, larger unit cells, and antipolar cation orderings. These intermediate phases are strongly reminiscent of -- when not identical to -- the centrosymmetric $Pbam$ crystal structure of the model antiferroelectric perovskite PbZrO$_3$, and are therefore very often described as antiferroelectric, even though actual antiferroelectric switching, i.e. the transition induced by an electric field, is difficult to observe due to the generally leaky character of BFO-based compounds and has been reported only in few studies~\cite{kan_composition_2011,mundy_liberating_2022}.

In this paper, we investigate the optical properties of rare-earth substituted BiFeO$_3$ systems at the ferroelectric-paraelectric phase transition. We focus on samarium-substituted BiFeO$_3$ (SmBFO) as a model system. Samarium has an ionic radius significantly smaller than bismuth (\SI{1.079}{\angstrom} vs. \SI{1.17}{\angstrom}), so the chemical pressure effect is quite drastic. Structural studies on powders~\cite{Pakalniskis2021,Pakalniskis2022} and ceramics~\cite{Arnold2015,Shi2016,Yu2018} report some degree of phase coexistence between $x\approx\SI{8}{\%}$ and $x\approx\SI{20}{\%}$, with a fraction of intermediate PbZrO$_3$-like phase peaking at $x\approx\SI{15}{\%}$. It is also the composition where AFE-like double hysteresis loops have been reported in thin films~\cite{kan_composition_2011,Cheng2009}. Overall, it appears that the details of the phases in the morphotropic region are complex, and their stabilization is sensitive to the processing route and the history of the samples. Optical studies at room temperature have shown that Sm-substitution causes a shift of the optical band gap toward lower energies in the polar phase and across the morphotropic region~\cite{Arora2014,Anthonyraj2015,Hu2017,Irfan2017,Kebede2020,Gu2021,Orudzhev2022}. The importance of the phase transition in this evolution however is unclear, and very different values have been reported. For example, Orudzhev {\it et al.} found a difference of \SI{100}{\meV}~\cite{Orudzhev2022} from pure BFO $R3c$ structure to the Sm 20\% $Pnma$ structure while Kebede {\it et al.} found a difference of \SI{410}{\meV}~\cite{Kebede2020}.

The purpose of this paper is to investigate specifically the role of the structural phase transition in Sm-substituted BFO on the optical properties. The motivation is to clarify the role of the phase transition as a tuning mechanism. This is relevant for tuning the optical properties by controlling the phase fractions in the morphotropic region. It is also important to evaluate the possibility to tune the optical absorption by inducing the phase transition with an electric field for antiferroelectric compositions, i.e. to evaluate the electrochromic properties of SmBFO. To do so, we performed measurements of optical properties as a function of temperature for several compositions across the morphotropic region. We demonstrate that this approach enables us to estimate the role of the structural transition in the change in the optical band gap. The origin of this change is discussed.  


\section*{Experimental details}

Powders of Sm-substituted BiFeO$_3$ (Sm$_x$Bi$_{1-x}$FeO$_3$ with $x=10$, 12, 14, 16, 18 and \SI{20}{\%}) were processed starting from the high quality oxides Sm$_2$O$_3$ from ChemPur, Bi$_2$O$_3$ form Alfa Aesar and Fe$_2$O$_3$ from Ventron by mechanochemical synthesis using a ball milling process. We used the PM100 miller from Verder. The starting oxides were mixed in a \SI{125}{m\liter} ZrO$_2$ jar with \SI{100}{\mm} ZrO$_2$ grinding balls, then rotated for \SI{24}{h}, alternating between \SI{2}{\min} rotation and \SI{2}{\min} break, resulting in effective rotation of \SI{8}{h} and \SI{8}{h} of break.\par

For each composition, powders and pellets have been processed. The powders were annealed at \SI{850}{\celsius}. Pellets of \SI{8}{\mm} in diameter were pressed with a uniaxial pressure of \SI{150}{\MPa}. The pellets were then sintered at \SI{850}{\celsius} at an atmospheric pressure of 1/2Al$_2$O$_3$+1/2Bi$_2$O$_3$ using a temperature ramp of \SI{350}{\celsius /h} for \SI{6}{h}. The phase purity and sample quality have been checked at room temperature by X-ray diffraction (XRD) on a D2 Bruker diffractometer using Cu K$_\alpha$ radiation and by Raman spectroscopy with a LabRam system from Horiba with a \SI{633}{\nm} laser, a laser power of \SI{1}{\mW} and a laser spot size of \SI{1.2}{\um} in diameter. The particle distribution and morphology was studied by a SEM field-emission gun Hitachi SU-70. All compositions show a faceted morphology and an average particle size of $\approx \SI{400}{\nm}$ (see Fig.~S1 in the supplementary information).\par

Raman and XRD measurements at room temperature overall confirm results known from the literature with a pure $R3c$ phase for low Sm concentrations (10 and 12\%) and a pure $Pnma$ phase for high Sm concentrations (20\%). Phase coexistence is observed for the intermediate Sm concentrations (14, 16 and 18\%); see Fig.~S2 in the supplementary information. Some very small peaks were occasionally observed that could indicate the presence of either the intermediate $Pbam$ phase or possibly a spurious phase, but in such small amounts that they do not affect the average provided by optical measurements. Besides, the Raman spectra did not give any evidence for any other phase than the $R3c$ phase and the $Pnma$ phase (Fig.~S3 in the supplementary information), in spite of the general sensitivity of Raman spectroscopy. As a result, in the following, we will discuss the data considering only the $R3c$ phase, the $Pnma$ phase, and a mixture of both. 

The optical properties were investigated by diffuse reflectance on powders in the UV-visible range using a PerkinElmer Lambda 850 equipped with a Harrick Praying Mantis diffuse reflectance accessory. The diffuse reflectance was measured across the phase transition for increasing- and decreasing-temperature runs up to \SI{800}{\K} for all Sm compositions. A BaSO$_4$ powder without absorption in the spectral range of interest was used as a reference for the total reflectance. The focused beam spot size was about \SI{2}{\square\mm}.

\par
We analyze the diffuse reflectance using Kubelka-Munk theory, neglecting specular reflectance~\cite{alcaraz_de_la_osa_extended_2019}. The Kubelka-Munk theory is a two-flux model with differential equation formalism using empirical coefficients to describe the scattering and absorption of light in a continuous medium. The Kubelka-Munk function $F(R_\infty)$ is given by the following equation with $R_\infty$ being the diffuse reflectance, $\alpha^{'}$ the Kubelka-Munk absorption constant, and $s$ the Kubelka-Munk scattering constant.\cite{lindberg_absolute_1987, simmons_diffuse_1975, simmons_reflectance_1976}
\[
F(R_\infty)=\frac{\alpha^{'}}{s}=\frac{(1-R_\infty)^2}{2R_\infty}\approx\alpha
\]

The Kubelka-Munk formalism assumes an infinitely thick sample and $s$ to be independent of the wavelength of the incident light. In this case $F(R_\infty)$ is proportional to the absorption coefficient $\alpha$.\cite{simmons_reflectance_1976}


\section*{Results and discussion}

\subsection*{Diffuse reflectance spectra at room temperature}

\begin{figure}[thbp]
    \centering
    \includegraphics[width=0.45\textwidth]{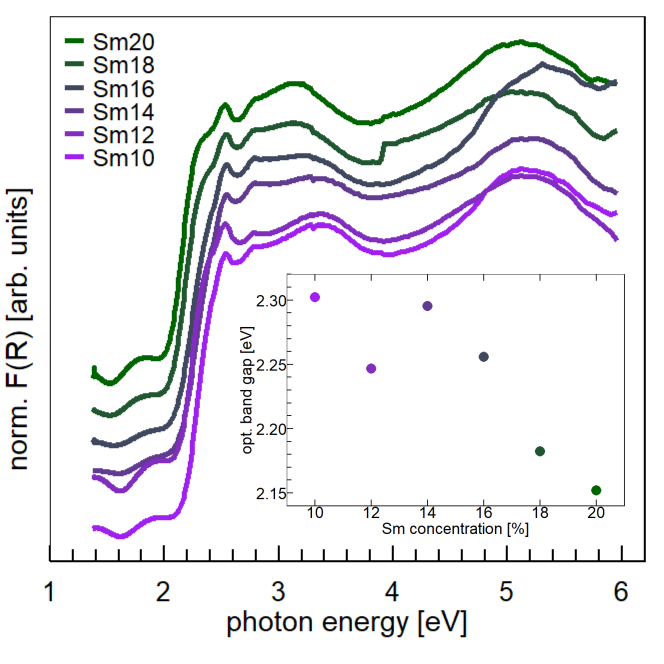}
    \vspace{-10 pt}
    \caption{Room temperature absorption spectra derived from the reflectance spectra using the Kubelka-Munk formalism. Spectra are normalized to the absorption peak at \SI{2.5}{\eV} slightly above the optical band gap and stacked to facilitate better comparison. The inset shows the decreasing general trend of the optical band gap energy with increasing Samarium concentration.}
    \label{fig:Op_RT}
\end{figure}

The diffuse reflectance spectra for all compositions are qualitatively comparable to those previously described for pure BFO powders~\cite{bai_size_2016}. Fig.~\ref{fig:Op_RT} shows the absorption spectra for the different compositions derived from the Kubelka-Munk formalism from the diffuse reflectance measurements presented in Fig.~S4 in the supplementary information. For comparison purposes, each absorption spectrum is normalized to the absorption peak at \SI{2.5}{\eV} above the band gap. The absorption spectra show several features. The most remarkable is the sharp and sudden increase around \SI{2}{\eV}, a typical feature of a band-to-band transition. This is associated to the band gap of the material.  At lower energies, the function exhibits weak bumps around \SI{1.5}{\eV} and \SI{1.9}{\eV} that are associated with Fe$^{3+}$ $d$--$d$ excitations.\cite{bai_size_2016,meggle_temperature-dependent_2019, xu_optical_2009}. At higher energies, broad absorption bands are visible at \SI{3.4}{\eV} and \SI{5.2}{\eV} and a narrow band at \SI{2.5}{\eV} that can be associated with higher electronic transitions.\cite{bai_size_2016}

With such a complex absorption profile, it is not obvious how to define the band gap. The presence of important absorption tails below the band gap precludes the use of classical Tauc plots. We choose here to define the optical band gap as the inflection point in the absorption edge as presented in Ref.~\cite{hill_role_2020}. In this way, we are not sensitive to the details of the absorption spectrum, including the direct or indirect band gap nature of the material, and do not depend on an arbitrary choice for a linear extrapolation in a Tauc plot. This also provides us with a way to treat consistently the temperature-dependent data. The inset of Fig.~\ref{fig:Op_RT} shows the optical band gap as a function of the Sm concentration. The band gap energy decreases with increasing Sm concentration, as observed in \cite{Arora2014, Hu2017, Kebede2020, Orudzhev2022}.\par

For Sm concentration of 10\%, we found a band gap energy of \SI{2.3}{\eV}. In the literature, the band gap energies for the same composition ranged from \SI{2.06}{\eV}~\cite{Hu2017, Kebede2020} to \SI{2.26}{\eV}~\cite{Arora2014}, with one study reporting a value of \SI{2.13}{\eV}~\cite{Orudzhev2022}. In all of these studies, measurements were performed on nanoparticles for which a strong size dependence on the optical properties is known~\cite{bai_size_2016} and the band gap energy was extracted using a Tauc plot. The large variations in the band gap energy from one work to the other result from variations in particle size and the challenge of consistently determining the band gap energy through the Tauc plot.\par
In the next step, we aim to investigate the temperature dependence of both absorption and the optical band gap. 

\subsection*{Optical band gap as a function of temperature}
All powders were measured by diffuse reflectance at high temperatures, both increasing- and decreasing-temperature runs being recorded. As a representative example, we show the absorption spectra derived from the diffuse reflectance measured at the increasing-temperature run for the composition of $x=14$\,\% in Fig.~\ref{fig:KMFunction}. The absorption spectra for all other compositions are shown in Fig.~S5 in the supplementary information. The absorption edge, indicated by a sudden increase in absorption, is found around \SI{2}{\eV} and strongly shifts with temperature, as expected from studies on BFO single crystals~\cite{palai__2008, weber_temperature_2016}. The weak absorption features below the absorption edge seem to disappear with increasing temperatures, see inset of Fig.~\ref{fig:KMFunction}. Qualitatively similar observations were made for all compositions.\par

\begin{figure}[th]
    \centering
    \includegraphics[width=0.45\textwidth]{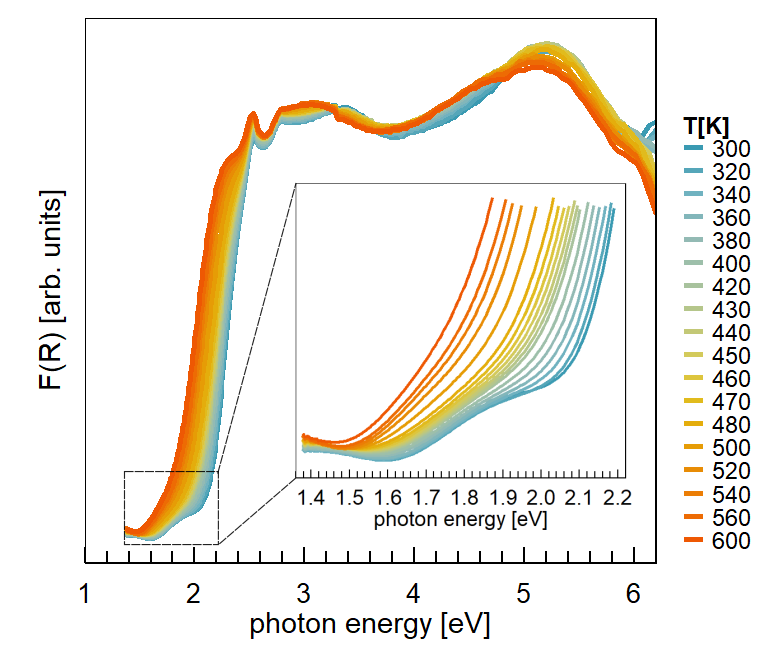}
    \caption{Absorption spectra derived from the diffuse reflectance measured at the increasing temperature run for the powder (Bi$_{0.86}$Sm$_{0.14}$)FeO$_3$ using by Kubelka-Munk functions $F(R)$. The inset shows a zoom on the low energy tails.}
    \label{fig:KMFunction}
\end{figure}

\begin{figure}[htbp]
    \centering
    \includegraphics[width=0.45\textwidth]{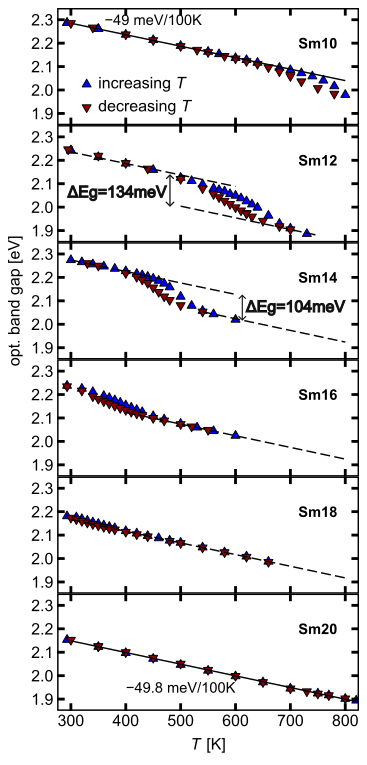}
    \caption{Optical band gap defined as the inflection point in the absorption edge of the Kubelka-Munk function $F(R)$ versus temperature for (Bi$_{1-x}$Sm$_{x}$)FeO$_3$ for x varying from \SI{10}{\%} to \SI{20}{\%} upon increasing and decreasing temperature runs. Solid lines are linear fits to the data. For the fits indicated by dashed lines the slope from low (high) temperatures and low (high) Sm concentrations was carried over.}
    \label{fig:absorptionedge}
\end{figure}

For every run, we extracted the optical band gap, i.e. the inflection point in the absorption edge, from the Kubelka-Munk functions $F(R)$ of all compositions for both the increasing and decreasing temperature runs. The results are shown in Fig.~\ref{fig:absorptionedge}. The blue triangles pointing up show the optical band gap from the increasing-temperature run, while the brown triangles pointing down represent the optical band gap from the decreasing-temperature run. 

For the two highest concentrations of Sm 18\% and 20\%, the optical band gap energy decreases linearly with temperature and does not show any anomaly, meaning there is no indication for a structural phase transition. \par

In all other cases, deviations from linearity are clearly visible. The most representative case is obtained for 14\% that shows two linear behaviors, one at low temperatures and one at high temperatures. In between, the band gap shows an inflection that we associated with the phase transition between the low-temperature polar $R3c$ phase and the high-temperature non-polar $Pnma$ phase. This change occurs smoothly over a wide temperature range of $\approx \SI{150}{\K}$. There is a very clear hysteresis between the increasing and decreasing temperature runs. The transition temperature, given by the inflection in the optical band gap, is $\approx \SI{50}{\K}$ lower for the decreasing temperature run than for the increasing temperature run. The presence of an hysteresis and the coexistence of phases are consistent with the first-order character of the rhombohedral to orthorhombic transition.\par

The other compositions show the same scenario, whereby the transition is shifted towards lower temperatures for higher Sm concentrations. For the highest concentrations of 18\% and 20\% Sm, no transition is observable within the measured temperature range; it would be expected to occur below room temperature. For 10\% Sm, the transition is only partially observed. Although we observe the onset of the transition and some hysteretic behavior, the maximum temperature of \SI{800}{\K}is not sufficient to reach the high-temperature linear behavior. The hysteresis is of practical importance for the preparation of samples at room temperature in the \qtyrange[range-phrase=~--~]{16}{18}{\%} range, that will depend strongly on the thermal history of the sample.\par

We fitted the linear regions of the band gap for Sm concentrations of 10\% and 20\%. The fits are shown as solid lines. Interestingly, comparing the slopes of the low-temperature and the high-temperature phase, the value is nearly identical with $\approx -\SI{0.49}{\meV\per\K}$. For the intermediate compositions where the linear regime is not as broad, we fixed the slope, using the slope of the fit for 10\% Sm for low temperatures and the slopes of the fits for 20\% Sm for high temperatures. These fits are represented by dashed lines. We believe, it is reasonable to assume that the slope must be identical from one composition to another, provided that the dominant phase is the same.\par

The value $-\SI{0.49}{\meV\per\K}$ is two to three times slower than for pure BFO with $\approx -\SI{1.3}{\meV\per\K}$~\cite{palai__2008}. However, when measured at the inflexion point where the slope is maximum and the change mostly driven by the phase transition, it reaches $-\SI{1.7}{\meV\per\K}$ and $-\SI{2.05}{\meV\per\K}$ for 12\% and 14\% respectively, which highlights the strong contribution of the phase transition in this evolution. For 12\% and 14\% Sm, we calculated the difference between the low-temperature and high-temperature linear fits, as a way to eliminate the effect of temperature and thermal expansion and isolate the contribution of the structural phase change. This results in values of $\Delta E_g$ of \SI{134}{\meV} and \SI{104}{\meV}, respectively. The smaller value for 14\% Sm is attributed to the fact that the phase transition is not complete and the sample still displays some degree of phase coexistence even at room temperature, as seen in XRD and discussed in the next section. We therefore retain 12\% Sm as the sample that is closest to a pure $R3c$ to pure $Pnma$ transition in the investigated temperature range, and \SI{134}{\meV} as the difference in optical band gap attributed to this phase transition. 

This value closely aligns with the variation in the optical band gap of ~\SI{200}{\meV} from rhombohedral to supertetragonal found in strained BFO thin films~\cite{Sando2016}. Considering that the engineered strain in this example is much larger than the strain introduced by the phase transition in our material system, it is reasonable to assume that the change in the optical band gap is directly proportional to the strain. We also note that our value is very close to the \SI{130}{\meV} reported for the comparison between the paraelectric cubic phase and the ferroelectric tetragonal phase in PbTiO$_3$.~\cite{Zelezny2016}.

\subsection*{Phase diagram}

\begin{figure}[htbp]
    \centering
    \includegraphics[width=0.5\textwidth]{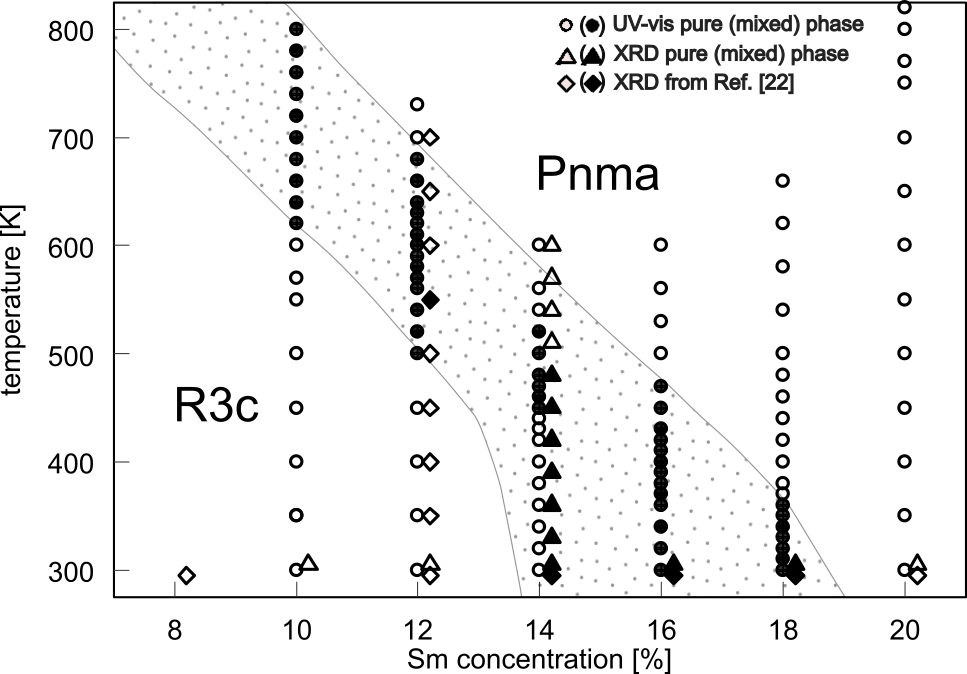}
    \caption{Phase diagram constructed using optical data (circles) and XRD data (triangles), supplemented by XRD data from literature~\cite{Pakalniskis2022} (diamonds). For clarity, the XRD data points are slightly shifted in respect to the exact Sm concentration. Open symbols indicate a pure phase ($R3c$ or $Pnma$) and filled symbols are data points where phase coexistence is observed.}
    \label{fig:phasediagram}
\end{figure}

Finally, we tentatively construct a phase diagram from the optical data, as shown in Fig.~\ref{fig:phasediagram}. We assume a pure phase where the optical band gap behaves linearly with temperature and a mixture of phases at temperatures where we observe a significant deviation ($\pm$ \SI{5}{\meV}) from linearity. The diagram is constructed using data collected upon increasing temperature. We can distinctly identify three regions: the polar $R3c$ phase at low temperatures and low Sm concentrations, the non-polar $Pnma$ phase at high temperatures and high Sm concentrations, and a mixture of both phases in between.\par

The so constructed phase diagram can be complemented by previously reported XRD measurements~\cite{Pakalniskis2022}, as well as our own XRD data at room temperature. Both the literature and our XRD data confirm the presence of a pure $R3c$ phase at room temperature for low Sm concentrations ranging from 8\% to 12\%. A pure $Pnma$ phase is found for the highest Sm concentration of 20\%. Near the phase transition, we observe the coexistence of the two phases. 14\% appears as a limiting case where phase coexistence is not immediately apparent from the optical data but is seen in XRD at room temperature. For clarification, we also carried out complementary XRD measurements as a function of temperature up to \SI{800}{\K} with increasing- and decreasing-temperature runs; see Fig.~S6 in the supplementary information. This experiment confirms the coexistence of both phases at room temperature, including after cooling down. The presence of phase coexistence is also consistent with the smaller $\Delta E_g$ measured for 14\% as compared to 12\%.

Overall, both optical and XRD data exhibit the same trends, but with differences in the range where phase coexistence is observed. In the optical data, the range for phase coexistence appears significantly broader in temperature. For 12\% Sm concentration, as temperature increases, literature reports a phase transition accompanied by phase coexistence at \SI{550}{\K}. This could also originate from the different length scales probed by the two techniques. While XRD gives a volume average, diffuse reflectance is a highly surface sensitive technique. This, combined with the famous skin effect of BFO and BFO-based materials, may lead to differences observed in phase coexistence and apparent transition temperatures.\par

\section*{Conclusion}
In summary, this paper shows an optical study of BiFeO$_3$ powders modified by Sm substitution with a concentration ranging from 10 to 20\% across the morphotropic region. We have shown that the optical band gap extracted by diffuse reflectance and its treatment via the Kubelka-Munk function reveal the expected phase transition between the polar BFO-like phase and the non-polar orthorhombic phase at high Sm concentration and high temperature. The phase diagram established from optical data appears to be in qualitative agreement with the known behavior of these solid solutions, but it is also characterized by a very large hysteresis of the transition temperature. We estimated that the phase transition comes with a shift in the band gap of about \SI{130}{\meV}, which we primarily attribute to the difference in volume strain between the two phases. In the context of a possible antiferroelectric behavior of SmBFO, we anticipate that this value gives a good estimation of the bandgap variation as the non-polar to polar transition is induced by an electric field, i.e. a good estimation of its potential electrochromic behavior.

\section*{Acknowledgments}
CH acknowledges funding from the Fond National de la Recherche under Project PRIDE/15/10935404. CMB acknowledges funding from the Fond National de la Recherche under Project BIAFET C16/MS/1134912/Guennou. BD acknowledges funding from the Fond National de la Recherche under Project INTER/MOBILITY/19/13992074 and the French National Research Agency (ANR) through CERACOOL ANR-23-CE05-0012, SOFIANE ANR-23-CE09-0007 and PHOTOTRICS ANR-24-CE08-0954-03. 

\section*{Conflicts of interest}
The authors have no conflicts to disclose.

\section*{Data availability}
The data that support the findings of this study are openly available in Zenodo at 10.5281/zenodo.15175369.

\section*{Supplementary Material}
See the supplementary information for: I.~Grain size and morphology (SEM analysis); II.~XRD patterns at room temperature; III.~Composition-dependent Raman analysis; IV.~Composition-dependent diffuse reflectance measurements; V.~Absorption as a function of temperature for different compositions; VI.~Temperature-dependent XRD analysis for Sm concentration of 14\%.

\clearpage
\bibliography{BFObiblio}

\end{document}